\documentstyle[epsfig,12pt]{article}
\newlength{\dinwidth}
\newlength{\dinmargin}
\newcommand{\ra}{\rightarrow}
\newcommand{\BGAMAXS}{B \ra X _{s} + \gamma}

\newcommand{\ba}{\begin{array}}
\newcommand{\ea}{\end{array}}
\newcommand{\be}{\begin{equation}}
\newcommand{\ee}{\end{equation}}
\newcommand{\bea}{\begin{eqnarray}}
\newcommand{\eea}{\end{eqnarray}}

\def\bra{\langle}
\def\ket{\rangle}

\def\a{\alpha}
\def\b{\beta}
\def\g{\gamma}

\def\e{\epsilon}
\def\p{\pi}

\def\l{\lambda}
\def\m{\mu}
\def\n{\nu}
\def\G{\Gamma}

\def\to{\rightarrow}
\begin{document}
\thispagestyle{empty}
\addtocounter{page}{-1}
\begin{flushright}
SLAC-PUB-96-7113\\
ZU-TH 2/1996\\
hep-ph/9602281\\
February 1996
\end{flushright}
\vspace*{3cm}
\centerline{\Large\bf 
Virtual Corrections to the Decay
$b \to s +\gamma$ \footnote{Partially supported by Schweizerischer
Nationalfonds.}}
\vspace*{2.0cm}
\centerline{\large\bf Christoph Greub }
\vspace*{0.4cm}
\centerline{\large\it SLAC Theory Division, Stanford University}
\centerline{\large\it Stanford, California 94309,,USA}
\vspace*{0.8cm}
\centerline{\large\bf Tobias Hurth \footnote{ address after March 1996: 
ITP, SUNY at Stony Brook, Stony Brook NY 11794-3840, USA} and  
Daniel Wyler}
\vspace*{0.4cm}
\centerline{\large\it Institute for Theoretical 
Physics, University of Z\"urich}
\centerline{\large\it Winterthurerstr. 190, CH-8057 Z\"urich, Switzerland}
\vspace*{0.5cm}
\vspace*{3cm}
\centerline{\Large\bf Abstract}
\vspace*{1cm}
We calculate the $O(\a_s)$ virtual corrections to the
matrix element for $b \to s \gamma$, taking
into account the
contributions of the four-Fermi operator
$O_2$ and the electromagnetic and color dipole-type
operators. The results are combined with 
existing $O(\a_s)$
Bremsstrahlung corrections 
in order to obtain
the relevant inclusive rate. The new
result 
drastically reduces the large scale 
dependence of the
leading logarithmic approximation.
It implies that
a very accurate prediction for the branching ratio 
for $B \to X_s \g
$ will become possible once also the corrections to the Wilson
coefficients are available.

\newpage
\section{Introduction}
\label{sec:introd}

In the Standard model (SM) flavor-changing neutral currents
only arise at the one-loop level. The corresponding B-meson decays
are therefore particularly
sensitive to 'new physics'; but also
within the Standard model framework, they can
be used to constrain several Cabibbo-Kobayashi-Maskawa matrix
elements involving the top-quark. For both reasons
precise experimental and theoretical work is of great importance.

$B \to K^* \gamma$ is the first rare B decay mode, which
has been measured in 1993 by the CLEO collaboration \cite{CLEOrare1}
and  recently also the first measurement of
the inclusive photon energy
spectrum and the branching ratio
in the decay $\BGAMAXS$ was reported \cite{CLEOrare2}.
In contrast to the exclusive channels, the inclusive mode allows a less
model-dependent comparison with theory, because no specific
bound state model is needed for the final state.

This data is in a good
agreement with
the SM-based theoretical computations presented
in \cite{agalt,aglett,shifmangamma}
given that large uncertainties exist in both
experimental and theoretical results.
In particular, the measured branching ratio $BR(B \to X_s \gamma)
= (2.32 \pm 0.67)
\times 10^{-4}$ \cite{CLEOrare2} overlaps with the
SM-based estimates in
\cite{agalt,aglett} and in \cite{Buras94,Ciuchini94}.

As the experiments are becoming more precise in the near future,
also the calculations must be refined in order to
allow quantitative statements about new
physics or standard model
parameters.

It is well known that
QCD corrections to the decay rate for $b \to s \g$
bring in
large logarithms of the form $\a_s^n(m_W) \, \log^m(m_b/M)$,
where $M=m_t$ or $m_W$ and $m \le n$ (with $n=0,1,2,...$).
These large terms can be resummed by renormalization
group techniques.
At present, only the leading logarithmic
corrections (i.e. $m=n$) have been calculated systematically.
In this work we include one class of next-to-leading
corrections which we describe in more detail below.

The calculations are most easily done in the
framework of an effective theory which is obtained
by integrating out the top quark and the $W$-boson
in the standard model and other heavy particles
in extensions thereof.
A complete set of dimension-6 operators relevant for the process
$b \to s \gamma$  (and $b \to s \g g$)
is contained in the effective Hamiltonian
\cite{Grinstein90}
\begin{equation}
\label{heff}
H_{eff}(b \to s \gamma)
       = - \frac{4 G_{F}}{\sqrt{2}} \, \lambda_{t} \, \sum_{j=1}^{8}
C_{j}(\mu) \, O_j(\mu) \quad ,
\end{equation}
where
$G_F$ is the Fermi constant
coupling constant and
$C_{j}(\mu) $ are the Wilson coefficients evaluated at the scale $\mu$,
and $\lambda_t=V_{tb}V_{ts}^*$ with $V_{ij}$ being the
CKM matrix elements.
The operators $O_j$ read
\bea
\label{operators}
O_1 &=& \left( \bar{c}_{L \b} \g^\m b_{L \a} \right) \,
        \left( \bar{s}_{L \a} \g_\m c_{L \b} \right)\,, \nonumber \\
O_2 &=& \left( \bar{c}_{L \a} \g^\m b_{L \a} \right) \,
        \left( \bar{s}_{L \b} \g_\m c_{L \b} \right) \,,\nonumber \\
O_7 &=& (e/16\p^{2}) \, \bar{s}_{\a} \, \sigma^{\m \n}
      \, (m_{b}(\mu)  R + m_{s}(\mu)  L) \, b_{\a} \ F_{\m \n} \,,
        \nonumber \\
O_8 &=& (g_s/16\p^{2}) \, \bar{s}_{\a} \, \sigma^{\m \n}
      \, (m_{b}(\mu)  R + m_{s}(\mu)  L) \, (\l^A_{\a \b}/2) \,b_{\b}
      \ G^A_{\m \n} \quad ,
        \nonumber \\
\eea
where $e$ and $g_s$ are the electromagnetic and the strong
coupling constants, respectively. In the magnetic moment type
operators $O_7$ and $O_8$, $F_{\m \n}$ and $G^A_{\m \n}$
are the electromagnetic and the gluonic field strength
tensors, respectively and
$L=(1-\g_5)/2$ and $R=(1+\g_5)/2$
stand for the left and right-handed projection operators.
We note here that the explicit mass factors in $O_7$
and $O_8$ are the running quark masses.
We
did not give explicitly
the four-Fermi operators $O_3$--$O_6$ in
eq. (\ref{operators}), because they have small
and negligible Wilson coefficients.

To leading logarithmic precision,
 it is consistent to perform the
matching of the effective and full theory
without taking into account QCD-corrections
\cite{Inami}
and to calculate
the anomalous dimension matrix (8 $\times$ 8)
to order $\a_s$ \cite{Ciuchini}.
The corresponding leading logarithmic Wilson coefficients
are given explicitly in
\cite{Buras94,AGM94}.
The leading logarithmic contribution to the
decay matrix element is then obtained by calculating the
tree-level matrix element of the operator $C_7 O_7$ and
the one-loop matrix elements of the four-Fermi operators
$C_i O_i$ ($i=1,...,6$).
In the NDR scheme which we will use in this paper,
the latter are non-zero only for $i=5,6$. Their effect
can be absorbed into a redefinition of $C_7 \to C_7^{eff}$
\footnote{For the analogous $b \to s g$ transition, the effects
of the four-Fermi operators can be absorbed by the shift
$C_8 \to C_8^{eff}=C_8 + C_5$.}
\be
\label{C78eff}
C_7^{eff} \equiv  C_7 + Q_d \, C_5 + 3 Q_d \, C_6 \quad .
\ee

Since the first order calculations  contain large
scale uncertainties,
it is important to take into account the next-to-leading
order corrections.
A complete next-to-leading calculation contains
two classes of improvements: First,
the Wilson coefficients are required to next-leading
order at the scale $\mu \approx m_b$. This requires
the matching
with the full theory (at $\mu=m_W$)
at the $O(\a_s)$ level and the renormalization
group equation has to be solved using the anomalous dimension
matrix calculated up to order $\a_s^2$.
Second,
the real and virtual $O(\a_s)$ corrections for the matrix element (at scale
$\mu \approx m_b$)
must be evaluated.

The higher order matching has been calculated in ref. \cite{adel}
and work on the Wilson coefficients is in progress.
In the present paper we complete
the second step.
While the Bremsstrahlung corrections have been worked out 
\cite{agalt,aglett,aglong,Pott}
in order to
get a non-trivial photon energy spectrum at the partonic level
for $B \to X_s \g$, the virtual corrections to $b \to s \g$
have not been completely known so far. Only those
virtual diagrams  which are needed to
cancel the infrared sigularities generated by the Bremsstrahlung
diagram were calculated.
In the present paper we
evaluate all the additional virtual
correction, neglecting, as mentioned the contributions of the small
operators
\footnote{
This omission will be a source of a slight scheme and scale dependence
of the next-to-leading order result.} $O_3$--$O_6$.
These new contributions
substantially
reduce the strong scale dependence of the leading calculation.

In the following we thus consider the virtual $O(\a_s)$
corrections to $b \to s \g$ due to 
the four-Fermi operator $O_2$
and the
magnetic operators $O_7$ (which has already been calculated in the
literature) and $O_8$ (which is new); note
that the operator $O_1$ does not contribute
to the matrix elements for $b \to s \g$ and $b \to s \g g$
because of its color structure.
As the corrections to $O_7$ and $O_8$ are one-loop diagrams,
they are relatively easy to work out.
In contrast, the corrections to $O_2$, involve two-loop
diagrams.

\section{Virtual Corrections to $O_2$, $O_7$ and $O_8$}

\begin{figure}[htb]
\vspace{0.10in}
\centerline{
\epsfig{file=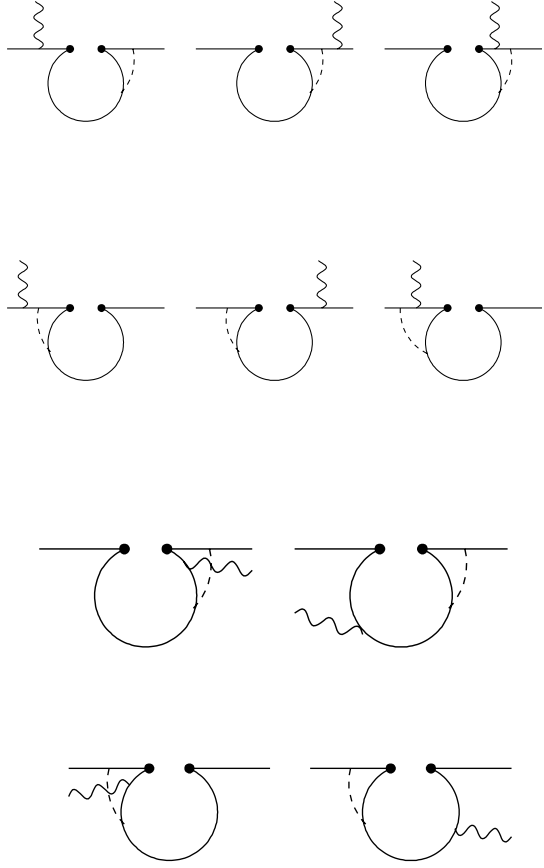,height=5in,angle=0}
}
\vspace{0.08in}
\caption[]{Non-vanishing two-loop diagrams 
associated with the operator $O_2$.
The fermions ($b$, $s$ and $c$ quark) are represented by solid lines.
The wavy (dashed) line represents the photon (gluon).
\label{fig:1}}
\end{figure}

Since the virtual and Bremsstrahlung
corrections to the matrix elements are only one (well-defined)
part of the whole next-to-leading program,
one expects that this contribution alone will depend on
the renormalization scheme used. Even within the modified minimal
subtraction scheme $(\overline{MS})$ used here,
two different
``prescriptions'' how to treat $\gamma_5$, will lead to different
answers. Since previous calculations of the Bremsstrahlung
diagrams have been done in the NDR scheme where also the leading
logarithmic Wilson coefficients are available,
we also use it here. A discussion of the results in the 
't Hooft-Veltman scheme (HV) \cite{HV}
is presented in ref. \cite{GHWlong}.
While the one-loop
$(\a_s^0)$  matrix element of the operator $O_2$ 
vanishes, we must consider
several two-loop contributions $(\a_s^1)$ shown
in Fig. \ref{fig:1}.
The diagrams in Fig. \ref{fig:1} are grouped in such a way that 
the sum 
of each line is gauge invariant.
The corresponding two-loop integrals are calculated by  standard 
Feynman parameter technique. 
The heart of our procedure
which we explain in detail in ref. \cite{GHWlong}
is to
represent the rather complicated denominators in the remaining Feynman 
parameter integrals as complex Mellin-Barnes integrals
\cite{Boos,Usyukina,Smirnov,Erdelyi}.
After inserting this
representation and interchanging the order of integration, the
Feynman parameter integrals are reduced to well-known Euler
beta-functions. Finally, the residue theorem
allows  to represent the remaining
complex integal as the sum over the residues taken at
the pole positions of
certain beta- and
gamma-functions; this naturally leads
to an expansion in the ratio $z=(m_c/m_b)^2$, which
numerically is about $z=0.1$.
It is, however, not a Taylor series
because it also involves logarithms of $z$.
A generic diagram
which we denote by $D$ has the form
\be
\label{generic}
D = c_0 + \sum_{n,m} c_{nm} z^n
\log^m z \quad ,  \quad z = \frac{m_c^2}{m_b^2} \quad ,
\ee
where the coefficients $c_0$ and $c_{n m}$ are independent
of $z$.
The power $n$ in eq. (\ref{generic})
is in general a natural multiple of $1/2$
and $m$ is a natural number including 0. In the explicit
calculation, the lowest $n$ turns out to be $n=1$.
This implies the important fact
 that the limit $m_c \to 0$ exists;
thus, there cannot be large logarithms (if the
charm quark is replaced by the up quark with
its small mass) in these diagrams.
As we show in \cite{GHWlong}, the power
$m$ of the logarithm is bounded by  $4$
independently of the value of $n$.
We also  proved that the expansion converges,
at least for $z \le 1/4$.
In our results, we have retained all terms up to $n=3$.
The numerical result truncated at
$n=3$ differs from the result for
$n=2$ by only about $1\%$.

The final result for the 
dimensionally regularized matrix element $M_2$
of the operator $O_2$
which represents the sum of all two-loop diagrams in 
\ref{fig:1} is 
(The individual sets of diagrams are given
in ref. \cite{GHWlong}.):
\bea
\label{m2res}
M_2 &=& \left\{ - \frac{23}{108 \e} \,
\left( \frac{m_b}{\mu} \right)^{-4\e} \,
+ \frac{1}{648} \left[ -833 +144\p^2 z^{3/2} \right. \right. \nonumber \\
&& \hspace{0.3cm}
+ \left( 36 L^3 +108 L^2+(-324 \p^2 +1296) \, L
-1296 \zeta(3) -180\p^2 +1728 \right) \, z \nonumber \\
&& \hspace{0.3cm}
+ \left( 36 L^3 +(-216 \p^2 +432) L +
648 +72\p^2 \right) \, z^2 \nonumber \\
&& \hspace{0.3cm} \left.
+ \left( 1092 L -756 L^2 - 84\p^2 -54
\right) z^3 \right] \nonumber \\
&& \hspace{0.3cm}
+ \frac{i \pi}{27} \left[ -5
+ \left( -3 \p^2 +45 +9 L^2 +9L \right) \, z \right. \nonumber \\
&& \hspace{0.3cm} \left. \left.
+(-3\p^2+9 L^2) z^2 + (-12 L +28) z^3 \right] \right\}
\nonumber \\
&& \hspace{0.3cm} \times
\frac{\a_s}{\p} \, C_F  \bra s \g |O_7| b \ket _{tree}
\eea
 Note that  
we introduced the renormalization scale  in the form
$\mu^2 \exp(\gamma_E) /(4 \pi)$ which
is convenient for $\overline{MS}$ subtraction.
Here, $z=(m_c/m_b)^2$ and $L = \log(z)$ and
$\e$ is the ultraviolet regulator in the convention, where 
$d=4-2\e$.
The symbol $\zeta$ in eq. (\ref{m2res}) denotes the Riemann
Zeta function, with $\zeta(3) \approx 1.2021$; $C_F=4/3$ is
a color factor and
the matrix element $\bra s \g |O_7| b \ket _{tree}$ is the
$O(\a_s^0)$ tree level
matrix element of the operator $O_7$. In eq. (\ref{m2res})
we have inserted the numerical values for the charges of
the up and down type quarks ($Q_u=2/3$ and $Q_d=-1/3$). 

There are also counterterms to be included.
As we are interested in 
contributions to $b \to s \g$ which are proportional to
$C_2$, we have to take, in addition to the two-loop matrix elements
of $C_2 O_2$, also the one-loop matrix elements of the four Fermi
operators
$C_2 \delta Z_{2j} O_j$ ($j=1,...,6$) and the tree level contribution 
of the magnetic operator $C_2 \delta Z_{27} O_7$.
In the NDR scheme  the
only non-vanishing contributions
to $b \to s \g$ come from  $j=5,6,7$.
The operator renormalization
constants $Z_{ij}$
can be extracted from the literature \cite{Ciuchini}
in the context of the leading
order anomalous dimension matrix:
\bea
\label{zfactors}
\delta Z_{25} &=&  -\frac{\a_s}{48 \pi \e} \, C_F \quad , \quad
\delta Z_{26} =   \frac{\a_s}{16 \pi \e} \, C_F \quad ,
\nonumber \\
\delta Z_{27} &=&  \frac{\a_s}{16 \pi \e} \,
(6 Q_u - \frac{8}{9} Q_d) \, C_F \quad .
\eea
Defining
\be
M_{2j} = \bra s \g |\delta Z_{2j} O_j|b \ket \quad ,
\ee
we find the following contributions to the matrix elements
\bea
\label{counter}
M_{25} &=& -\frac{\a_s}{48 \pi} \, Q_d C_F \,
  \frac{1}{\e} \left(
\frac{m_b}{\mu} \right)^{-2\e}
\, \bra s \g |O_7| b \ket _{tree} \nonumber \\
M_{26} &=& \frac{3 \a_s}{16 \pi} \, Q_d C_F \,
  \frac{1}{\e} \left(
\frac{m_b}{\mu} \right)^{-2\e}
\, \bra s \g |O_7| b \ket _{tree} \nonumber \\
M_{27} &=& \frac{\a_s}{ \pi} \,
\left( \frac{3 Q_u C_F}{8} - \frac{Q_d C_F}{18} \right)
\, \frac{1}{\e} \, \bra s \g |O_7| b \ket _{tree}
\eea
We note that there is no one-loop contribution to $b \to s \g$
from the counterterm proportional to $(C_2 \frac{1}{\e} O_{12}^{ev})$,
where the evanescent operator $O_{12}^{ev}$
(see
e.g. the last ref. in \cite{Ciuchini}) reads 
\be
\label{evop}
O_{12}^{ev} = \frac{1}{6} O_{2} \left( \g_\m 
\to \g_{[\m}\g_\n \g_{\rho ]} \right) - O_2 \quad .
\ee
Adding the two-loop expression $M_2$ (eq.(\ref{m2res}))
and the counterterms 
(eq. (\ref{counter})), we find the renormalized result, which
can be written as
\be
\label{m2ren}
M_2^{ren} = \bra s \g |O_7| b \ket _{tree} \, \frac{\a_s}{4 \p} \,
\left( \ell_2 \log \frac{m_b}{\mu}  + r_2 \right) \quad ,
\ee
with
\be
\label{l2}
\ell_2 = \frac{416}{81}.
\ee
\bea
\label{rer2ndr}
\Re r_2 &=& \frac{2}{243} \, \left\{- 833 + 144 \pi^2 z^{3/2}
\right. \nonumber \\
&& \hspace{0.3cm}
+ \left[ 1728 -180 \pi^2 -1296 \zeta (3) + (1296-324 \pi^2) L +
108 L^2 + 36 L^3 \right] \, z \nonumber \\
&& \hspace{0.3cm}
+ \left[ 648 + 72 \pi^2 + (432 - 216 \pi^2) L + 36 L^3 \right] \, z^2
\nonumber \\
&& \hspace{0.3cm}        \left.                 +
\left[ -54 - 84 \pi^2 + 1092 L - 756 L^2 \right] \, z^3 \, \right\}
\eea
\bea
\label{imr2ndr}
\Im r_2 &=& \frac{16 \p}{81} \, \left\{- 5
+ \left[ 45-3 \pi^2 + 9 L +
9 L^2 \right] \, z
+ \left[ -3 \pi^2 + 9 L^2 \right] \, z^2 +
\left[ 28 - 12 L  \right] \, z^3 \, \right\}
\eea
Here, $\Re r_2$ and $\Im r_2$ denote the real and the imaginary part
of $r_2$, respectively.

The virtual corrections associated with the
operator $O_7$
have been taken into account
by Ali and Greub, see e.g. \cite{agalt,aglett,aglong}.
These corrections contain infrared singularities and for
$m_s=0$ also collinear singularities.
At the level of the decay rate, these singularities
cancel if one adds the Bremsstahlung correction based
on the matrix element squared of the operator $O_7$.
Denoting by  $\G_{77}$ the sum of
the lowest order contribution, the virtual
$O(\a_s)$ correction and those parts of the Bremsstrahlung correction
just mentioned,
this finite quantity
reads
in the limit
$m_s=0$
\be
\label{Gamma7}
\G_{77} = \G_{77}^0 \,
\left[ 1 + \frac{\a_s}{3 \p} \left(
\frac{16}{3} - \frac{4\p^2}{3} + 4 \log \frac{m_b}{\mu} \right)
\right]  \quad ,
\ee
where  the lowest order contribution $\G_{77}^0$ is given by
\be
\label{Gamma70}
\G_{77}^0 = \frac{m_b^2(\mu) m_b^3}{32 \p^4} \,
|G_F \l_t C_7^{eff}|^2 \,
\a_{em} \quad .
\ee

For later convenience,
it is useful to define a 'modified' matrix element for
$b \to s \g$,
in such a way that its square reproduces
the result in eq. (\ref{Gamma7}). This modified matrix element
$M_7^{mod}$ reads
\be
\label{m7lr}
M_7^{mod} = \bra s \g|O_7| b \ket _{tree}
\, \left( 1+ \frac{\a_s}{4\p} \left( \ell_7 \, \log \frac{m_b}{\mu}
+r_7
\right) \right)
\ee
with
\be
\label{l7r7}
\ell_7 = \frac{8}{3}  \quad , \quad
r_7 = \frac{8}{9} \, (4 - \pi^2) \quad .
\ee

\begin{figure}[htb]
\vspace{0.10in}
\centerline{
\epsfig{file=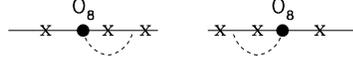,height=2in,angle=-90}
}
\vspace{0.08in}
\caption[]{Contributions of $O_8$ to $b \to s \g$. The cross (x)
denotes the possible place where the photon is emitted. 
\label{fig:2}}
\end{figure}

Finally, we consider the contributions to $b \to s \g$
generated by the operator $O_8$, i. e., the matrix element
\be
\label{m8def}
M_8 = \bra s \g|O_8|b \ket \quad.
\ee
The corresponding Feynman diagrams
are shown in Fig. \ref{fig:2}.
The sum of all 6 diagrams in Fig.\ref{fig:2} yields
\be
\label{ress123456}
M_8 = \frac{Q_d \, C_F}{12} \, \frac{\a_s}{\p} \,
\left[ -\frac{12}{\e} -33 +2\p^2 +24 \log(m_b/\mu) -6 i \p
\right]  \,
\bra s \g|O_7|b \ket _{tree}
\quad .
\ee
There is also a contribution from a counterterm,
reading
\be
\label{m8counter}
M_8(counter) = \delta Z_{87} \,
\bra s \g|O_7|b \ket _{tree}   \quad ,
\ee
where the renormalization constant
\be
\delta Z_{87} =  \frac{\a_s}{\p} \, C_F Q_d \,\frac{1}{\e}
\ee
can again be
taken from the literature \cite{Ciuchini}.
We thus arrive at the
renormalized result $M_8$:
\be
\label{m8lr}
M_8^{ren} = \bra s \g |O_7| b \ket _{tree} \, \frac{\a_s}{4 \p} \,
\left( \ell_8 \log \frac{m_b}{\mu}  + r_8 \right) \quad ,
\ee
with
\be
\label{l8r8}
\ell_8 = - \frac{32}{9} \quad , \quad
r_8 = - \frac{4}{27} \,
\left( -33 +2\p^2  -6 i \p
\right)
\quad .
\ee
\section{Impact on the branching ratio}
To summarize, we have calculated the virtual
corrections to $b \to s \gamma$ coming from the operators
$O_2$, $O_7$ and $O_8$. The contributions from the other
operators in eq. (\ref{heff}) are either small or vanish.
As discussed above, some of the Bremsstrahlung
corrections to  the operator $O_7$ have been transferred into
the matrix element $M_7^{mod}$ for $b \to s \gamma$
in order to simplify the following discussion.
While we have neglected
those virtual corrections of the operators $O_3$-$O_6$
which are given by the analogous diagrams as shown in Fig.
\ref{fig:1},
we took into account the non-vanishing
diagrams of  $O_5$ and $O_6$
where the gluon connects the external
quark lines and the photon is radiated from the charm quark;
these corrections are automatically taken into account
when using $C_7^{eff}$ instead of $C_7$.
Since the Wilson coefficients of the omitted
operators are about fifty times smaller than that
of the leading one and since we expect that their matrix
elements can be enhanced at most by color 
factors, it seems reasonable to neglect them.

We can now easily
write down the amplitude $A(b \to s \g)$
for $b \to s \gamma$
by summing the various contributions derived in the previous
section. We follow closely the treatment of
Buras et al. \cite{Buras94}, where the general structure
of the next-to-leading order result is discussed in detail.
We write
\be
\label{amplitudevirt}
A(b \to s \g) = -\frac{4 G_F \l_t}{\sqrt{2}} \, \hat{D} \,
\bra s \g|O_7(\mu)|b \ket _{tree}
\ee
with
$\hat{D}$
\be
\label{dhat}
\hat{D} = C_7^{eff}(\mu) + \frac{\a_s(\mu)}{4\p} \left(
C_i^{(0)eff}(\mu) \ell_i \log \frac{m_b}{\mu} +
C_i^{(0)eff} r_i
\right)         \quad ,
\ee
and where the quantities $\ell_i$ and $r_i$ are given
for $i=2,7,8$ in eqs. (\ref{l2},\ref{rer2ndr}),
(\ref{l7r7}) and (\ref{l8r8}), respectively.
The first term, $C_7^{eff}(\mu)$,
on the rhs of eq. (\ref{dhat})  has to be
taken up to  next-to-leading logarithmic precision in order
to get the full next-to-leading logarithmic result, whereas
it is sufficient to use the leading logarithmic values of
the other Wilson coefficients in eq. (\ref{dhat}).
As the next-to-leading coefficient $C_7^{eff}$ is not known
yet, we replace it in the numerical evaluation by its
leading logarithmic value $C_7^{(0)eff}$.
The notation $\bra s \g|O_7(\mu)|b \ket _{tree}$ in
eq. (\ref{amplitudevirt}) indicates that the explicit $m_b$
factor in the operator $O_7$ is the running mass taken
at the scale $\mu$.

Since the relevant scale for a $b$ quark decay is expected to
be $\mu \sim m_b$, we expand the matrix elements of the
operators around
$\mu=m_b$ up to order $O(\a_s)$.
The result is
\be
\label{amplitudevirtuell}
A(b \to s \g) = -\frac{4 G_F \l_t}{\sqrt{2}} \, D \,
\bra s \g|O_7(m_b)|b \ket _{tree}
\ee
with
$D$
\be
\label{d}
D = C_7^{eff}(\mu) + \frac{\a_s(m_b)}{4\p} \left(
C_i^{(0)eff}(\mu) \gamma_{i7}^{(0)eff} \log \frac{m_b}{\mu} +
C_i^{(0)eff} r_i
\right)         \quad ,
\ee
where the quantities $\gamma_{i7}^{(0)eff}=\ell_i + 8 \, \delta_{i7}$
are just the entries of the (effective) leading order anomalous
dimension matrix \cite{Buras94}.
As also pointed out in this reference,
the explicit logarithms of the form
$\a_s(m_b) \log(m_b/\mu)$
in eq. (\ref{d})
are cancelled by the $\mu$-dependence of $C_7^{(0)eff}(\mu)$.
\footnote{As we neglect the virtual correction of $O_3$-$O_6$,
there is a small left-over
$\mu$ dependence.}
Therefore the scale dependence is
significantly reduced by including the virtual corrections
calculated in this paper.

The decay width $\G^{virt}$ which follows
from $A(b \to s \g)$ in eq.
(\ref{amplitudevirtuell}) reads
\be
\label{widthvirt}
\G^{virt} = \frac{m_{b,pole}^5 \, G_F^2 \l_t^2 \a_{em}}{32 \p^4}
\, F \, |D|^2 \quad ,
\ee
where in fact we discard the term of $O(\a_s^2)$
in $|D|^2$.
The factor $F$ in eq. (\ref{widthvirt}) is
\be
F = \left( \frac{m_b(\mu=m_b)}{m_{b,pole}} \right)^2 =
1- \frac{8}{3} \,  \frac{\a_s(m_b)}{\p} \quad .
\ee

To get the inclusive decay width for $b \to s \g (g)$, also
the Bremsstrahlung corrections (except the part
we have already absorbed) must be added.  
The contribution of the operators
$O_2$ and $O_7$ have been calculated before by
Ali and Greub \cite{agalt}, 
recently also
the complete set has been
worked out \cite{aglett,aglong,Pott}.
In the present work we neglect the small contribution
of the operators $O_3$ -- $O_6$ in analogy to the virtual corrections,
where only  $O_2$, $O_7$ and $O_8$ were considered also.

The branching ratio $\mbox{BR}(b \to s \g (g))$ is then obtained
by deviding the decay width $\G = \G^{virt} + \G^{brems}$
for $b \to s \g (g)$ by the theoretical expression for the
semileptonic width $\G_{sl}$ 
(including the well-known $O(\a_s)$ radiative
corrections \cite{CCM}) and by
multiplying with the measured semileptonic branching ratio
$\mbox{BR}_{sl} = (10.4 \pm 0.4)\%$ \cite{Gibbons}.

\begin{figure}[htb]
\vspace{0.10in}
\centerline{
\epsfig{file=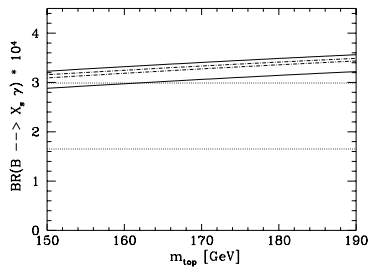,height=2in,angle=-90}
}
\vspace{0.08in}
\caption[]{Branching ratio for $b \to s \g (g)$
based on the formulae in this section. The different
curves are explained in the text.
\label{fig:3}}
\end{figure}

In Fig. \ref{fig:3} we present the
result for the branching ratio for $b \to s \g (g)$
based on the contributions discussed above.
A rather crucial parameter is the ratio $m_c/m_b$;
it enters both,
$b \to s \g$ mainly through the
virtual corrections of $O_2$ and
the semileptonic decay width through phase space.
To estimate this ratio we 
put $m_{b,pole}=4.8 \pm 0.15$ GeV for the $b$ quark pole mass;
{}from the mass difference $m_b - m_c=3.40$ GeV, which 
is known quite precisely through the $1/m_Q$
expansion \cite{HQET2,Shifmanetal}, 
one then gets $m_c/m_b=0.29 \pm 0.02$.
In the curves we have used the central values for $m_{b,pole}$ and
$m_c/m_b$. 
For the CKM matrix elements we put $V_{cb}=V_{ts}$ and $V_{tb}=1$.
In Fig. \ref{fig:3} we have plotted the calculated branching 
ratio as a function of the top quark mass $m_t$.
The horizontal dotted
curves show the CLEO limits for the branching ratio
$\mbox{BR}(B \to X_s \g)$ \cite{CLEOrare2}. 
To illustrate the scale dependence of the branching ratio,
we varied the scale $\mu$ between $(m_b/2)$ and $(2m_b)$.
We considered two  'scenarios' which differ by higher order
terms. First, we   
put the scale $\mu=m_b$ in the explicit
$\a_s$ factor in eq. (\ref{d}) and in the correction
to the semileptonic decay width, as it was also done
by Buras et al. \cite{Buras94}. The resulting $\mu$ dependence
is shown by dash-dotted lines.
Second, we retain the scale $\mu$ in the
explicit $\a_s$ factors; this leads to the solid curves in
Fig. \ref{fig:3}. In both cases the upper curve corresponds
to $\mu=m_b/2$ and the lower curve to $\mu=2m_b$. 
We mention that the $\mu$-band is larger in the second
scenario and it is therefore safer to use this band to illustrate
the remaining scale uncertainties. 

\begin{figure}[htb]
\vspace{0.10in}
\centerline{
\epsfig{file=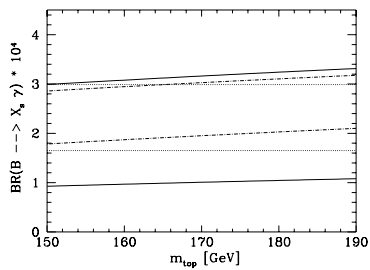,height=2in,angle=-90}
}
\vspace{0.08in}
\caption[]{Branching ratio for $b \to s \g(g)$.
The leading logarithmic result is shown by the dash-dotted curve;
The solid line shows the situation before the virtual contributions of
$O_2$ and $O_8$. See text.
\label{fig:4}}
\end{figure}

In Fig. \ref{fig:4} we show for comparison
the leading logarithmic
result for the branching ratio for $b \to s \g$,
based on the tree-level matrix element of the operator $O_7$
and using the tree-level formula for the semileptonic decay
width. Varying the scale $\mu$ in the same range as above, leads
to the dash-dotted curves in Fig. \ref{fig:4}. 
We have also plotted 
the result as is was available
before the inclusion of the virtual corrections of $O_2$ and
$O_8$ (but with Bremsstrahlung and virtual corrections to $O_7$
included). 
This is reproduced by putting
$\ell_2=r_2=\ell_8=r_8=0$ in our formulae. 
As noticed in the literature  \cite{agalt,Pott},
the $\mu$ dependence in this case (solid lines) is even larger
than in the leading
logarithmic result.
 
{}From the results in Fig. \ref{fig:4} 
it was relatively easy to read off
a reasonable
prediction for the branching ratio within a large error
which was essentially determined by the $\mu$ dependence.
In the improved calculation (Fig. \ref{fig:3}) the $\mu$
dependence is significantly reduced, because all the logarithms
of the form $\a_s(m_b) \log(m_b/\mu)$ are cancelled as discussed
above. In the present situation it is, however,
premature to extract a prediction
for the branching ratio from Fig. \ref{fig:3}.
This only will be possible
when also  $C_7^{eff}$  is known up to next-to-leading
logarithmic precision. But this result will, essentially,
shift the narrow bands of Fig. \ref{fig:3}, without
broadening them significantly. Thus, a very precise
prediction will become possible and renewed experimental
efforts will be required. It is, however,
 rewarding to see 
that the next-to-leading result will lead to 
a strongly improved 
determination of the standard model parameters
or to better limits to new physics.

\vspace*{2cm}

{\bf Acknowledgements}
Discussions with A. Ali, S. Brodsky, M. Lautenbacher, M. Peskin
and L. Reina are thankfully acknowledged.
We are particularly indebted to M. Misiak
for many useful comments;  especially
his remarks concerning the
renormalization scale dependence were extremely useful.
One of us (C.G.) would like to thank the
Institute for Theoretical Physics in Z\"urich
for the kind hospitality.


\end{document}